\documentclass[12pt]{iopart}
\usepackage{graphicx}
\usepackage{dcolumn}
\usepackage{epsf}
\usepackage{bm}
\usepackage{times}

\def\eqref#1{(\ref{#1})}
\usepackage{graphicx}
\usepackage{harvard}

\newcommand{\addrGaithersburg}{National Institute of Standards and Technology, 
Gaithersburg, MD 20899, USA}
\newcommand{\addrMissouri}{Missouri University of Science and
Technology, Rolla, MO 65409, USA}
\begin{document}

\title[Fundamental constants and tests of theory]
{Fundamental constants and tests of theory in Rydberg states of one-electron ions}

\author{Ulrich D~Jentschura$^1$, Peter J~Mohr$^2$ and Joseph N~Tan$^2$}
\address{$^1$~\addrMissouri\\ $^2$~\addrGaithersburg}

\eads{\mailto{jentschurau@mst.edu}, \mailto{mohr@nist.gov}, \mailto{jtan@nist.gov}}

\date{\today}

\begin{abstract}

The nature of the theory of circular Rydberg states of
hydrogenlike ions allows highly-accurate predictions to be made
for energy levels.  In particular, uncertainties arising from
the problematic nuclear size correction which beset low
angular-momentum states are negligibly small for the high
angular-momentum states. The largest remaining source of
uncertainty can be addressed with the help of quantum
electrodynamics (QED) calculations, including a new
nonperturbative result reported here. More stringent tests of
theory and an improved determination of the Rydberg constant may
be possible if predictions can be compared with precision
frequency measurements in this regime.  The diversity of
information can be increased by utilizing a variety of
combinations of ions and Ryberg states to determine fundamental
constants and test theory.

\end{abstract}
\submitto{\JPB}
\maketitle
%
%%%%%%%%%%%%%%%%%%%%%%%%%%%%%%%%%%%%%%%%%%%%%%%%%%%%%%%%%%%%%%%%%%%%%

\section{Introduction}

Quantum electrodynamics (QED), the quantum field theory of
electrons and photons, is the first satisfactory quantum
description of the interaction of charged particles (and
antiparticles) via the exchange of photons and of the creation
and annihilation of elementary particles.   QED makes precise
predictions of various physical quantities which have been
tested across a vast array of phenomena.  Spectroscopic
measurements in atoms, however, have played a crucial role in
spurring the development of QED to be the most accurate physical
theory yet invented. With the quantization of the
electromagnetic field, QED does not remove the divergences well
known in Maxwell's classical theory of electromagnetism; on the
contrary, new infinities are found in QED, associated with
virtual processes of the vacuum.  Sensible, finite results are
obtainable only after renormalization; this is an art
inextricably tied to the introduction of fundamental constants,
such as the electron mass and charge, into the theory. 

The first tests of the emerging formalism of QED came shortly
after World War II.  \citeasnoun{1947001} reported the
first measurement of the anomalous magnetic moment of the
electron; in the same year, \citeasnoun{1947003} presented the
first measurement of the ``Lamb shift'' in the 2$s$ level of
hydrogen \cite{1951011}, another departure from the Dirac theory
of the hydrogen atom.  These discoveries have led to more
stringent tests of QED, with remarkable progress over six
decades. On the one hand, with control and miniminization of
cavity effects that limited early geonium $|g|-2$ experiments
\cite{1987003,1986027}, the magnetic moment of the electron $|g|
= 2(1 + a_{\rm e})$ has been measured recently at Harvard
University with a relative uncertainty of $2.8\times10^{-13}$
using a single electron isolated in a cylindrical Penning trap
\cite{2008043}; by comparison, the calculated magnetic moment of
the electron has a relative uncertainty of $5.2\times10^{-12}$
coming entirely from the uncertainty of the best independent
determination of the fine-structure constant.  On the other
hand, the hydrogen $1s-2s$ transition has been measured with a
relative uncertainty of $1.4\times10^{-14}$ \cite{2005082}; here
a test of theory is hampered by uncertainties from nuclear size
corrections, which currently limit the accuracy in the
determination of the Rydberg constant to a relative uncertainty
of $6.6\times10^{-12}$ \cite{2008145}.  It seems astonishing
that QED should have attained such high accuracy in the
abstractions employed for representing physical objects and
measurements, particularly when one of its pioneers has noted
the ``mathematical inconsistencies and renormalized infinities
swept under the rug'' \cite{dyson}.

Looking to the future, in this paper, we consider possible
determinations of fundamental constants and tests of theory in
circular Rydberg states of one-electron ions.  This has been
discussed by \citeasnoun{2008089}, where it was pointed out that
the problems that limit the theoretical predictions in low
angular-momentum states are strongly supressed in circular
Rydberg states, because the electron has a very small
probability of being near the nucleus in such states.  However,
even though some problematic aspects of the theory are absent in
these states, there is still a contribution from QED that needs
to be accounted for in order to make accurate predictions for
the energy levels.  That contribution was estimated by
calculating the largest unknown term in the expansion of the
electron self-energy in powers of $Z\alpha$.  It was then argued
that, based on the general trends of calculated terms in the
series, the value of the unknown term $A_{60}$ should be
sufficient to provide an accurate result, and the value of this
coefficient for various states relevant to the experiments of
interest was calculated.  Here we revisit the theory of the
energy levels and give results of an all-orders calculation that
confirms the expectation that the $A_{60}$ term provides an
accurate result.

\section{Theory}
\label{thr}

In principle, it is known that the $Z\alpha$ expansion converges
very well for Rydberg states. However, if one of the most
important physical constants is to be determined on the basis of
spectroscopic measurements---the Rydberg constant---then one
would like to have very reliable predictions.  The recent
proposal by \citeasnoun{2008089} considered, in particular, precise measurements
for transitions among Rydberg states of hydrogenlike ions with
medium charge numbers $Z = 10, \dots, 20$ and in the range of
principal quantum numbers $n=10, \dots, 20$, with unit change in
the principal quantum number $n$.  These transitions, according to
Fig.~1 of \citeasnoun{2008089}, are in the optical (THz) domain.

\begin{table}[thb]
\caption{\label{table1} Values of $A_{60}$ used by \citeasnoun{2008089} 
for states with principal quantum numbers $n=13,14,15$.}
\begin{indented}
\item[]
\begin{tabular}{crc}
\br
$n$ & $\kappa$ & $A_{60}$ \\
\mr
13 &$  11 $& 0.000\,006\,795\,75(5) \\
13 &$ -12 $& 0.000\,043\,189\,98(5) \\
13 &$  12 $& 0.000\,004\,699\,73(5) \\
13 &$ -13 $& 0.000\,027\,294\,75(5) \\
14 &$  12 $& 0.000\,004\,108\,25(5) \\
14 &$ -13 $& 0.000\,029\,799\,37(5) \\
14 &$  13 $& 0.000\,002\,966\,41(5) \\
14 &$ -14 $& 0.000\,019\,452\,79(5) \\
15 &$  13 $& 0.000\,002\,521\,08(5) \\
15 &$ -14 $& 0.000\,021\,160\,50(5) \\
15 &$  14 $& 0.000\,001\,893\,09(5) \\
15 &$ -15 $& 0.000\,014\,206\,31(5) \\
\br
\end{tabular}
\end{indented}
\end{table}

In \citeasnoun{2008089}, the self-energy remainder 
function is represented as 
\begin{eqnarray}
G(Z\alpha) &=& A_{60} + A_{81}(Z\alpha)^2\ln{(Z\alpha)^{-2}}
+ A_{80}(Z\alpha)^2 + \dots
\nonumber\\
&& + \frac{\alpha}{\pi}\,B_{60}  + \dots
+ \left(\frac{\alpha}{\pi}\right)^2\,C_{60} + \dots
\label{eq:gval}
\end{eqnarray}
where the $A$, $B$, and $C$ coefficients refer to one-loop,
two-loop, and three-loop effects, respectively.  This expansion
is valid for states with angular momentum $l \geq 2$. The main
contribution to $G$ comes from the one-loop self-energy, and
indeed, for states with angular momentum $l \geq 3$, the
coefficients $A_{81}$ and $A_{80}$ are determined exclusively by
the self-energy, because the vacuum-polarization contribution to
these coefficients vanishes. For even higher angular momenta ($l
\geq 10$), we can estimate vacuum polarization effects to be
entirely negligible at a relative accuracy level of $10^{-15}$
for the transitions under consideration.  We therefore define
the one-loop self-energy (SE) remainder function as
\begin{eqnarray}
\label{Zdependence}
G_{\rm SE}(Z\alpha) &=& A_{60} + A_{81}(Z\alpha)^2\ln{(Z\alpha)^{-2}}
+ A_{80}(Z\alpha)^2 + \dots
\end{eqnarray}
where the coefficients are understood to be entirely due to
self-energy effects. In \citeasnoun{2008089}, the following
approximation was made,
\begin{equation}
\label{approx}
G_{\rm SE}(Z\alpha) \approx A_{60} \,,
\end{equation}
and $A_{60}$ was calculated using a semi-analytic method.  The
$A_{60}$ coefficient depends only on the principal quantum
number $n$ of the state under investigation, and on the Dirac
angular momentum quantum number $\kappa$. Because the $A_{81}$
and $A_{80}$ coefficients can be expected to be numerically
small (this is in line with the trend exhibited by lower-order
coefficients as a function of $n$ and $\kappa$), the
approximation in Eq.~\eqref{approx} can be used with good effect even
for the range $10 \leq Z \leq 20$, but on the other hand, it
may appear questionable that all the $Z$-dependent terms
from Eq.~\eqref{Zdependence} are ignored in formulating the 
approximation~\eqref{approx}.  Values of $A_{60}$ used by
\citeasnoun{2008089} are summarized in Table~\ref{table1}, for
states with principal quantum numbers $n=13,14,15$.

\begin{table}[thb]
\caption{\label{table2} Values of the nonperturbative
self-energy remainder function $G_{\rm SE}(Z\alpha)$ for
near-circular Rydberg states with principal quantum numbers
$n=13,14,15$, and for nuclear charge numbers $Z =14,16$.}
\begin{indented}
\item[]
\begin{tabular}{crcccrcc}
\br
$n$ & $\kappa$ & $Z$ & $G_{\rm SE}(Z\alpha)$ & $n$ & 
$\kappa$ & $Z$ & $G_{\rm SE}(Z\alpha)$ \\
\mr
13 &$  11 $&14 &0.000\,006\,76(9) & 13 &$  11 $& 16 & 0.000\,006\,82(4)\\
13 &$ -12 $&14 &0.000\,043\,17(5) & 13 &$ -12 $& 16 & 0.000\,043\,21(2)\\
13 &$  12 $&14 &0.000\,004\,69(9) & 13 &$  12 $& 16 & 0.000\,004\,68(7)\\
13 &$ -13 $&14 &0.000\,027\,28(9) & 13 &$ -13 $& 16 & 0.000\,027\,28(5)\\
14 &$  12 $&14 &0.000\,004\,03(8) & 14 &$  12 $& 16 & 0.000\,004\,08(6)\\
14 &$ -13 $&14 &0.000\,029\,74(5) & 14 &$ -13 $& 16 & 0.000\,029\,78(3)\\
14 &$  13 $&14 &0.000\,002\,96(9) & 14 &$  13 $& 16 & 0.000\,002\,96(9)\\
14 &$ -14 $&14 &0.000\,019\,44(9) & 14 &$ -14 $& 16 & 0.000\,019\,45(9)\\
15 &$  13 $&14 &0.000\,002\,43(9) & 15 &$  13 $& 16 & 0.000\,002\,49(3)\\
15 &$ -14 $&14 &0.000\,021\,07(9) & 15 &$ -14 $& 16 & 0.000\,021\,14(2)\\
15 &$  14 $&14 &0.000\,001\,84(7) & 15 &$  14 $& 16 & 0.000\,001\,91(9)\\
15 &$ -15 $&14 &0.000\,014\,15(7) & 15 &$ -15 $& 16 & 0.000\,014\,23(9)\\
\br
\end{tabular}
\end{indented}
\end{table}

Recently, we have carried out a fully nonperturbative
calculation of QED self-energy shifts for Rydberg states. This
calculation incorporates the entire $Z$-dependence of the
right-hand side of~\eqref{approx} and takes into account all
higher-order (in $Z\alpha$) effects.  The basic method of
calculation is described by \citeasnoun{2001072}, but
siginificant refinements have been made for the case at hand.
Here, we discuss nonperturbative results in the range $n = 13,
14, 15$ of principal quantum numbers. Special emphasis will be
placed on the state with $n = 13$, $\kappa = 12$, and nuclear
charge numbers $Z = 14$ and $Z = 16$ [see also
Eqs.~\eqref{blog}---\eqref{a60} below]. In order to appreciate
the difficulty associated with this nonperturbative calculation,
we must recall a few facts about the bound-electron self-energy.
In general, the radiative energy shift is complex quantity whose
imaginary part corresponds to the one-photon decay width of the
reference state. Here, we are interested only in the real part
of the energy shift, which is the relevant physical quantity for
spectroscopic measurements, but this means that the pole terms
which are otherwise responsible for the nonvanishing imaginary
part, have to be subtracted.  

In the fully relativistic formalism, there are no selection
rules that would allow only dipole transitions to occur, and all
possible electric as well as magnetic multipoles have to be
taken into account.  If the reference state were the $2S$ state,
then, e.g., the integrand of the $2S$ self-energy would have a
pole due to an one-photon M1 transition to the ground state.
This would be the only pole to subtract for this calculation.
For the state with quantum numbers $n = 13$, $\kappa = 12$, a
total of 166 lower-lying bound-state poles have to be
subtracted. For the highest state considered here, which is the
state with $n = 15$, $\kappa = -14$, this number increases to
224. The calculation of the pole terms proceeds by a highly
accurate integration of angular variables, with a Green function
that is restricted to the virtual state in question. This
calculation yields the residue at the bound-state energy and is
subtracted before the final integration over the virtual photon
energy, with the full bound-electron Green function that
corresponds to a sum over all possible virtual bound states.
Extreme care must be taken in this subtraction operation, and
both the pole terms as well as the Green function term have to
be calculated to sufficient accuracy in order to retain
numerical significance in the results.  These subtractions
affect the so-called low-energy part of the calculation, where
the virtual photon energy is of the same order-of-magnitude as a
virtual atomic transition. 

There is a further ``trap'' to be avoided in the low-energy
part. Normally, for lower-lying states, one observes the
apparent convergence of terms in the Green function as more and
more partial waves are added and terminates the summation once
apparent convergence to a specified accuracy is reached.  This
method would have disastrous consequences for Rydberg states.
The reason is that for low-energy virtual photons, dipole
transitions from the reference state give rise to the
numerically dominant contribution to the self-energy integrand.
These correspond to virtual states with $|\kappa|$ being
displaced from the reference-state $\kappa$ Dirac quantum number
by no more than two. Therefore, the convergence criterion has to
be modified, and the summation over angular momenta of the
virtual states cannot be terminated before the reference-state
$\kappa$ is reached.

Further numerical problems plague the highly excited Rydberg
states in the so-called high-energy part, where virtual photons
of the order of the rest mass energy of the electron are
considered. In this region, it is customary to expand the
self-energy in powers of Coulomb interactions, i.e., in the
number of Coulomb photon exchanges of the electron with the
atomic nucleus. The so-called mass renormalization is associated
with the zero-Coulomb-vertex term, and the vertex
renormalization is associated with the one-Coulomb-vertex
diagram. It is especially problematic to evaluate the
subtraction terms associated with the zero-Coulomb-vertex and
the one-Coulomb-vertex diagrams for the Rydberg states. When
these calculations are performed in momentum space, numerical
cancellations occur because the Rydberg states are weakly bound
as compared to lower-lying states, and slow convergence of the
momentum space integrals is observed.  Surprisingly, there are
also quite severe numerical cancellations in the evaluation of
the fully relativistic momentum-space wave functions themselves.
Although their well-known expressions entail only a finite
number of terms and although these can be calculated recursively
\cite{1992009}, there are numerical cancellations among the
terms which cannot easily be avoided, and enhanced-precision
arithmetic has to be used.  In the evaluation of the high-energy
remainder terms, the weak binding of the Rydberg electron
implies that many partial waves have to be summed before
convergence is reached.  In a typical case, one has to sum about
one to two million $\kappa$ for the ground-state electron
self-energy in atomic hydrogen, and much less for higher nuclear
charge numbers. Here, an enhanced-accuracy relativistic Green
function is used, and about five million $\kappa$ are summed
before convergence is reached [see also \citeasnoun{convacc}].

We therefore use octuple (roughly 64-digit) precision for the
fully relativistic Green function of a hydrogenlike bound system
in order to describe the virtual excitations of the hydrogenlike
ion. The implementation of octuple-precision arithmetic is based
on a separation of the eightfold-precision number into
double-precision parts \cite{korobov}.  The enhanced accuracy is
used in order to overcome the convergence problems in angular
momentum expansions and the severe numerical losses due to
renormalization for weakly bound Rydberg states.  The
calculations were carried out on a cluster of IBM POWER5 64-bit
workstations at Missouri University of Science and
Technology.~\footnote{Certain commercial equipment, instruments,
or materials are identified in this paper to foster
understanding.  Such identification does not imply
recommendation or endorsement by the Missouri University of
Science and Technology nor by the National Institute of
Standards and Technology, nor does it imply that the materials
or equipment identified are necessarily the best available for
the purpose.}

After the subtraction of the lower-order analytic terms,
including the Bethe logarithm~\cite{2005013},
\begin{equation}
\label{blog}
\ln k_0(n =13, l = 12) = -0.315\,815\,189(1) \times 10^{-4} \,,
\end{equation}
we obtain 
\numparts
\begin{eqnarray}
\label{gse}
G_{\rm{SE}}(n = 13, \kappa = 12, Z = 14) &=& \; 4.69(9) \times 10^{-6} \,,
\\[2ex]
G_{\rm{SE}}(n = 13, \kappa = 12, Z = 16) &=& \; 4.68(7) \times 10^{-6} \,.
\end{eqnarray}
\endnumparts
These nonperturbative results have to be compared to the known
analytic result for the $A_{60}$ coefficient~\cite{2008089},
which is equivalent to the self-energy remainder at vanishing
nuclear charge $Z=0$,
\begin{equation}
\label{a60}
A_{60} =
G_{\rm{SE}}(n = 13, \kappa = 12, Z = 0) = \; 4.699\,73(5) \times 10^{-6}
\,.
\end{equation}
Within the numerical uncertainty, the results in Eq.~(\ref{gse})
are equal to those in Eq.~(\ref{a60}), and therefore, the
approximation indicated by Eq.~\eqref{approx} in estimating the
higher-order remainder function for the Rydberg states can be
justified {\em a posteriori}.  We reemphasize that the results
for $G_{\rm{SE}}$ from Eq.~\eqref{gse} contain the entire
higher-order nonperturbative remainder beyond $A_{60}$, and thus
the entire series starting with $A_{60}$, $A_{81}$, {\em etc.}
An analogous pattern can be observed for other transitions (see
Table~\ref{table2}).  Within the quoted numerical uncertainty of
the fully nonperturbative results for the remainder function,
which we estimated based on the apparent convergence of the
multi-dimensional integrals, the nonperturbative result for
$G_{\rm{SE}}$ is equal to the perturbative result for $A_{60}$.
This highlights the internal consistency of the theoretical
approach to the transitions in question.  Both the analytic as
well as the numerical calculations are highly nontrivial, and
mistakes therefore cannot be excluded.  However, the consistency
of two completely independent approaches is reassuring.

To summarize, we conclude that (a) the higher-order remainder
beyond $A_{61}$ is described to excellent accuracy by the
$A_{60}$ coefficient, even at medium nuclear charge numbers $Z
\geq 10$, and (b) the analytic coefficients multiplying the
higher-order corrections beyond $A_{60}$ are small, being
consistent with a general trend observed qualitatively by
\citeasnoun{2003108} and quantified by \citeasnoun{2008089}.  We
also observe that (c) there is mutual consistency of the
numerical and analytic approaches, which enhances the
reliability of the theoretical predictions.

\section{Experimental considerations}
\label{exp}

Rydberg states of hydrogenlike ions with $l \leq 2$ essentially
avoid a number of problems associated with either higher-order
binding corrections to QED interactions or the nuclear size
effect in Lamb shift predictions. Circular states have other
features which may be useful in experiments: They have the
longest lifetime in a given shell $n$ and a suppressed Stark
effect. In the cases being considered here, the higher-order QED
binding corrections for Rydberg states are smaller by a factor
of about $10^7$ compared to S states, providing significant
advantages from a theoretical point of view.  Such
simplifications come with experimental trade-offs associated
with a large spontaneous emission rate, for example.  Natural
decay linewidths tend to be small for states from which electric
dipole (E1) decay is forbidden (as in the case of the 2S level).
In contrast, the spontaneous decay rate for a circular Rydberg
state is dominated by an electric dipole E1 transition from the
highest-$l$ value of the state $\it{n}$ to the highest-$l$ value
of the state $\it{n-1}$.  The nonrelativistic expression for
this decay rate has been examined in \citeasnoun{BS} and also by
\citeasnoun{2005140} as the nonrelativistic limit of
theimaginary part of the level shift.  It is possible that this
decay process can introduce asymmetries into the lineshape for
the transition frequencies.  Fortunately, these effects are
small and of the order $\alpha(Z\alpha)^2E_{\rm QED}$ as has
been shown by \citeasnoun{1952012}.  These corrections, should
they become necessary, can be calculated and determined for the
particular systems chosen for the experiments, taking into
account details of the experimental set-up.  Precision
experiments with one-electron ions in Rydberg states 
require considerable effort to develop.
It is encouraging to note
that circular Rydberg states of hydrogen have been studied with
high precision in atomic beams for transitions in the millimeter
region; as a result, a determination of the Rydberg constant
with a relative uncertainty of $2.1 \times 10^{-11}$ has been
made \cite{th01dv}.  The advent of optical frequency combs
\cite{2006403} has opened the possibility of making precise
measurements of optical transitions between Rydberg states in
one-electron ions, with potential for higher precision
\cite{2004013} and broader range of applications \cite{2004312}.
Concurrently, an array of tools and techniques have emerged for
realizing ``engineered atoms'' that are built in traps and
tailored into states of interest. Cooling techniques
\cite{itano1995} developed for low-$Z$ ions, for instance, can
be extended to high-$Z$ ions extracted from sources like an
electron beam ion source/trap (EBIS/T) [See the review by
\citeasnoun{Donets1998}]. Some such synthetic atoms or
artificial quantum systems---for example, antihydrogen
\cite{2005236} or a single electron in a Penning trap
\cite{1986027}---have the potential to extend the range of
precision measurements that probe nature. 

Experiments with cold hydrogen-like ions in high-$l$ Rydberg
states may be possible for a wide range of nuclear charge $Z$
and angular momentum $l$ and would be useful for consistency
checks and optimization. An effort is underway at NIST to make
hydrogenlike ions starting from bare nuclei, with focus on
low-$Z$ to mid-$Z$ ions ($ 1 < Z < 11$).  Ion traps, such as a
compact Penning trap \cite{bat}, are being developed for
experiments to utilize bare nuclei extracted from the NIST EBIT
in its current configuration [see \citeasnoun{randr},
\citeasnoun{beamline}, \citeasnoun{nistebit}].  Low-$Z$
hydrogen-like ions, with their narrower line widths for optical
transitions among Rydberg states, seem the most favorable for
frequency comb-based measurements that could lead to better
determinations of fundamental constants---the Rydberg constant
in particular.  On the other hand, perturbations in Rydberg
states ({\it e.g.,} Stark mixing) tend to be significantly
attenuated in heavier hydrogen-like ions because a higher
nuclear charge $Z$ produces a larger fine-structure separation,
which scales as $Z^4$.  This wide range of available $(n,Z)$
combinations could be useful for extending diversity of
experiments used to determine fundamental constants and test
theory.  Apart from this long-term goal, experimental techniques
developed in this effort may directly enhance research in other
areas, such as: laboratory astrophysics, plasma diagnostics, new
regimes/techniques in atomic spectroscopy and standards
development.

\section{Conclusion}

In this paper (Sec.~\ref{thr}), we report on a nonperturbative
(in $Z\alpha$) calculation of the self-energy remainder function
for highly excited, nearly circular Rydberg states in
hydrogenlike ions with charge numbers $Z = 14$ and $Z = 16$ (see
Table~\ref{table2}).  These numerical values eliminate the
uncertainty due to uncalculated higher-order terms in the
expansion of the bound-electron self energy in $Z \alpha$, where
the growth of the number of terms in this expansion inhibits
analytic calculations.  The results of the nonperturbative
calculations of the self-energy for circular Rydberg states
reported here show that the approximation made by
\citeasnoun{2008089} in using the $A_{60}$ term in the
perturbation expansion to evaluate the energy levels is fully
justified. The situation is even better: Namely, the values for
the nonperturbative self-energy remainder listed in
Table~\ref{table2} agree with the perturbative $A_{60}$
coefficients listed in Table~\ref{table1} on the level of at
least $10^{-7}$ for $A_{60}$, even though the coupling constant
$Z\alpha$ is no longer negligible at the indicated values of the
nuclear charge number (we recall that it is the absolute, not
the relative, accuracy of the self-energy remainder function
that determines the predictive limits of the calculations).
This confirms the expectation that the remainder terms
drastically decrease for Rydberg states in comparison to
lower-lying ionic states.  We also confirm that accurate
evaluations of the energy levels of these states can be made,
with the potential to provide an independent value for the
Rydberg constant and an additional test of theory.

The aforementioned agreement provides additional evidence for
the Rydberg states being extremely ``favorable'' states with
regard to the theoretical analysis. Not only does the
self-energy function converge much more rapidly than for
lower-lying states [see Sec.~\ref{thr}
and~\citeasnoun{2008089}], but also, other problematic effects
(such as the nuclear-size correction to the energy levels) are
so vanishingly small for these states so as to be entirely
negligible at current and projected levels of accuracy.  In
particular, the nuclear-size correction does not need to be
taken into account in the prospective determination of the
Rydberg constant from measurements which are currently being
planned (see Sec.~\ref{exp}).

We conclude with a few more general remarks.  On the one hand,
the electron in a Rydberg state is bound sufficiently loosely to
avoid complications due to the shape of the nuclear charge
distribution and other problematic issues.  On the other hand,
it is bound sufficiently strongly so that its energy levels
provide additional insight into the nature of one of the most
precisely understood interactions in nature and allow one to
derive the Rydberg constant from a precise measurement using
ionic Rydberg states, which are appropriately named for this
purpose.  Advances both in experimental techniques (see
Sec.~\ref{exp}) as well as in the theoretical analysis
(Sec.~\ref{thr}) are indispensable along the way.

% Acknowledgments
\section*{Acknowledgments}

UDJ has been supported by the National Science Foundation (Grant
PHY--8555454) as well as by a Precision Measurement Grant from
the National Institute of Standards and Technology.

\end{document}